\documentclass[preprintnumbers,superscriptaddress,endnote,nofootinbib,prd]{revtex4}
\usepackage{graphicx}
\usepackage{graphicx}
\usepackage{bm}
\usepackage{epsfig}
\usepackage{amssymb}
\usepackage{amsmath}

\newcommand{\slsh}[1]{\not\!{#1}}
\newcommand{\be}{\begin{eqnarray}}
\newcommand{\ee}{\end{eqnarray}}
\newcommand{\bpm}{\begin{pmatrix}}
\newcommand{\epm}{\end{pmatrix}}
\newcommand{\CL}{\mathcal{L}}
\newcommand{\nn}{\nonumber}
\newcommand{\lsim}{\lesssim}
\newcommand{\gsim}{\gtrsim}
\def\kpirc{k\pi r_c }
\newcommand{\beq}{\begin{equation}}
\newcommand{\eeq}{\end{equation}}
\newcommand{\ord}[1]{\mathcal{O}{(#1)}}

\begin{document}

\begin{center}
{\Large \bf Natural Neutrino Masses and Mixings from Warped Geometry}
\vskip .17in
{\bf Gilad Perez}$^a$ and
{\bf Lisa Randall}$^b$
\vskip .18in
\end{center}
\vskip .04in
$^a $\textit{C.~N.~Yang Institute for Theoretical Physics,
State University of New York, Stony Brook, NY 11794-3840, USA.}
\\
$^{b} $\textit{Jefferson Laboratory of Physics, Harvard University
Cambridge, Massachusetts 02138, USA.}

\begin{center} {\bf Abstract}\\\end{center}

We demonstrate that flavor symmetries in warped geometry can provide a natural explanation for large mixing angles and economically explain the distinction between the quark and lepton flavor sectors. We show how to naturally generate Majorana neutrino masses assuming a gauged a $U(1)_{B-L}$ symmetry broken in the UV that generates see-saw masses of the right size. This model requires lepton minimal flavor violation (LMFV) in which only Yukawa matrices (present on the IR brane) break the flavor symmetries. 
The symmetry-breaking is transmitted to charged lepton bulk mass parameters as well to generate the hierarchy of charged lepton masses. With LMFV, a GIM-like mechanism prevents dangerous flavor-changing processes for charged leptons and permits flavor-changing processes only in the presence of the neutrino Yukawa interaction and are therefore suppressed when the overall scale for the neutrino Yukawa matrix is slightly smaller than one in units of the curvature. In this case the theory can be consistent with a cutoff of 10 TeV and 3 TeV Kaluza-Klein masses. 

 
\vskip .2in

\renewcommand{\thepage}{\arabic{page}}
\setcounter{page}{1}

\section{Introduction}

Understanding the flavor structure of the Standard Model (SM) presents one of the most daunting challenges to particle physicists. The masses range from the top quark at 170 GeV to the up quark of about 5 MeV to the electron mass of 500 KeV to neutrinos with masses fractions of an eV. If that is not bad enough, the mixing angles vary from a fraction of a percent in the quark sector to order unity in the lepton sector, posing another major challenge to model-builders. 

Clearly this spectrum is a great mystery. But viewed from another perspective, we have a large number of clues in the quark and lepton spectrum that might help us ultimately understand what lies beyond the Standard Model.
This is the attitude that has been taken toward the hierarchy problem that it is a clue to physics beyond the Standard Model.  It is worth noting that any theory that has been designed to address this problem has implications for flavor, unfortunately, usually bad ones that seriously constrain the model.

Neutrino masses and mixings should not be treated as separate problems from the hierarchy problem and quark masses and mixings. Ultimately we want a single framework where all of these parameters are as natural as possible.  Even without a specific new mechanism it is of interest to see if it all fits together.

In this paper we show how flavor might actually be explained in the context of warped geometry models that address the hierarchy problem \cite{RS1} and furthermore flavor violation can be suppressed to an acceptable level.
We now know  \cite{AS,RSfermions} that in extra-dimensional models with bulk fermions one can readily obtain flavor hierarchies due to the exponential dependence of the IR wavefunction value on the bulk fermion mass. 
Furthermore in models with an anarchic Yukawa structure--that is, all Yukawa entries are of the same order--mixing angles naturally align with quark masses since the IR fermion wavefunction values determine both.
Such models do surprisingly well with respect to flavor violations, given the assumed anarchic mixing structure and the light KK masses.

In Ref. \cite{FPR} we showed how to align the flavor structure of the bulk masses and Yukawas in models where only the Yukawa violates the flavor symmetry and prevent dangerous flavor changing effects. We now call this pseudo-minimal flavor violation 5D Quark Minimal Flavor Violation (QMFV). 

We show how to generate charged lepton and neutrino mass matrices in the context of warped extra dimensional models. We identify several novel features and insights into flavor associated with these models. We show  that  one can account for both the range of masses and the range of mixing angles quite readily in the context of a warped extra-dimensional flavor model with a pseudo-minimal flavor violation mechanism built in. The QMFV structure is applied to the lepton sector where 
 5D Lepton Minimal Flavor Violation (LMFV) is imposed. 5D LMFV implies that the lepton flavor group is only broken by the 5D leptonic Yukawa matrices. 
 Not only are the stringent flavor constraints satisfied but masses and mixing angles are explained.

The first key point is that Yukawas are anarchic in these models, where anarchic means that all entries are of order unity and no small are large numbers in fundamental Yukawas are assumed. 
Small mixing angles arise from ratios of wavefunctions in the IR so small mixing angles arise when wavefunctions are nondegenerate. 
That is precisely the problem of the split fermions mechanism when dealing with the neutrino flavor parameters which show an anarchical structure. 
With sufficiently degenerate bulk masses, wavefunctions are degenerate and mixing angles are big of the order of the size in the original anarchic Yukawa.

The second key point is that a gauged lepton number is required to prevent overly-large Majorana masses for the neutrinos generated in the IR. Without restrictive assumptions about lepton-symmetry-breaking in the UV this implies a see-saw generation for the light neutrinos where the Majorana mass for the right-handed neutrinos is generated in the UV and are therefore large.

 Given the size of charged lepton masses, the only way to suppress neutrino masses in a Dirac scenario is  to have the neutrino wavefunction be small in the IR and big in the UV. However, one needs to gauge lepton number and break this gauged lepton number (gauged due to AdS/CFT correspondence) in the UV (the UV to sequester from the IR and not generate overly large Majorana mass there). So to have a Dirac scenario one needs to break lepton number in such a way that it is not transferred to the right-handed neutrinos. This is possible, but generally requires a careful choice of charge for the lepton number breaking field and leaves a residual discrete symmetry.  Furthermore such a scenario would not necessarily explain large mixing angles and the necessary flavor suppression that must be present. 

We consider the alternative more generic possibility (from the point of view of lepton-number symmetry breaking). In this case, we are forced to have a Majorana mass for the right-handed neutrinos and a see-saw mechanism.  The advantage of placing this see-saw in a warped scenario   (or any extra-dimensional geometry) that one can naturally explain the discrepancy between the Majorana and Planck or GUT scales as well as the hierarchy of masses through exponential suppressions of bulk wavefunctions.

The remaining question is then to understand simultaneously the small mixing angles in the quark sector vs the big mixing angles in the lepton sector. We explain how in our warped context this naturally translates into a difference of mass splittings of order 10 percent vs of order 1 percent. The latter we say can alternatively be understood as the consequence of a flavor symmetry.

We mention the idea of see-saw and exponentials is not in and of itself new. What is new is the understanding of how to get large mixing angles. We generically take 5d Yukawas to be anarchic and then understand in terms of degenerate wavefunctions why these angles are carried over the the 4d Yukawas. 

The remaining issue is then how to simultaneously allow for large mixing angles and small flavor-changing effects. This is truly new in our paper and is a consequence in our models of the LMFV flavor symmetry.  Note that this symmetry only allows for moderate additional flavor suppression, so is natural to use only in the warped RS context. Otherwise flavor violations would be too big with a low KK scale.

Our final point is that flavor considerations in this context imply a pseudo-GIM-like structure  that we call LMFV (lepton minimal flavor violation) as with the quarks in which only Yukawas break the flavor symmetry. We will investigate this structure in more detail in the following section.

We find that large mixing angles, such as those in the neutrino sector,
can be readily accounted for in this framework, which assumes the most natural form for Yukawa matrices namely completely anarchical. We then show that with the LMFV assumption and the neutrino Yukawa of order 0.1 that lepton flavor is conserved to the desired degree of accuracy.

We also can understand why quarks and leptons have such different mixing structure in this framework. The difference between small mixing angles correlated with masses as in the quark sector and large mixing angles independent of masses as with neutrinos is entirely attributable to a number of order 0.1, namely 
the degree of breaking of the SU(3) flavor by the bulk masses, as
encoded in the Yukawa-dependent piece of the bulk mass terms
 in the quark and lepton sector.
Furthermore the range of masses is readily accounted for if we assume SU(3) degenerate bulk masses broken perturbatively (except maybe for the top quark) by Yukawas. If this model is right there are a few flavor diagonal parameters that 
vary between about one one-hundredth and order unity
that account for the mysterious spectrum of quark and lepton masses without generating overly-large flavor-changing effects.

 Other models in the literature have also discussed the neutrino masses in the context of warped geometry.
It was shown that the small neutrino masses can be naturally obtained both in the case of Dirac~\cite{RSfermions,DiracMajorana} and Majorana~\cite{hsMaj} neutrino masses.
attempted 
to explain lepton flavor structure in warped geometry. 
This is possible but in this case the gauged lepton symmetry that must be present must be broken in a nongeneric way. 
However there is generally no natural explanation for large mixing angles, and still getting a large enough tau mass (although see Ref. \cite{ramanlatest} for a possible explanation). Even in this case, without LMFV, there is a fairly strong constraint on the KK scale. 
We elaborate on  these points in the following section where we explain the role of gauged flavor symmetry.

\section{Gauged Flavor Symmetry}
 Here we focus on warped extra dimension models that address the hierarchy problem~\cite{RS1}.
In extra dimension models~\cite{RSfermions,AS} with bulk fermions, hierarchies in the flavor parameters are easily obtained.
Thus one can straightforwardly construct a model which naturally explain the hierarchical structure of the SM quark flavor sector~\cite{AS,hs}.
However, one is not guaranteed that the resulting model would be consistent with flavor constraints and in general this is not the case.

 In some respects the status of flavor precision tests in Randall-Sundrum (RS) models is similar to the one related 
to precision electroweak tests of the models, in particular the $T$ parameter.
In the SM the $T$ parameter is protected by a custodial global symmetry of the Higgs sector which is broken
only weakly by the hypercharge interaction and by the third generation Yukawa coupling (which is of order unity even for the top and therefore can be consider as weak).
The RS1 dual theory contains strongly coupled dynamics that need not respect those symmetries and therefore can induce large uncontrollable contributions to the $T$ parameter, especially due to the fact that the third generation quarks are composite states~\cite{APR,cust1}.
If, however,  the strong dynamics (or the composite sector) is invariant under a custodial symmetry, the corrections to the $T$
parameter are suppressed~\cite{cust1}.

Now let us discuss the flavor sector of the theory and the status of flavor precision tests.
Within the SM the  global flavor symmetries are weakly broken, mainly by the the third generation Yukawa couplings and by the neutrino masses. This implies highly suppressed contributions to processes involve flavor changing neutral currents (FCNCs), lepton family conversion and lepton number violation.
However, generically, the RS1 strong dynamics violates these symmetries~\cite{aps}.
For instance, the Yukawa interactions  between the fermion KK states and the fermion couplings to the gauge states are anarchic and
misaligned~\cite{aps}.
Interestingly, there is extra protection (compare to split fermion models in flat extra dimension~\cite{Flat}) against flavor violation
due to the fact that the light generations are mostly elementary and therefore have suppressed couplings to the 
flavor violating strong dynamics sector~\cite{aps,NMFV}, as well as due to the flat KK gluon wave function in the UV.
 However, even the presence of third generation flavor violation (NMFV or next-to-minimal-flavor-violation) is enough
to induce dangerous contributions to various flavor violating processes~\cite{hs,aps, Kitano:2000wr,abp,CFW}. 

This need not be the case if the strong dynamics possess the same global symmetries as the SM.
Precisely as in the case of the custodial symmetry, one can gauge the SM flavor symmetry in the bulk
in a way that the strong dynamics is flavor preserving. 
One can then have exact alignment between the sources of flavor violation and
also have the flavor breaking realized in a controllable manner.

In Ref. \cite{FPR} we showed how this can work for quarks with QMFV. 
According to this assumption the bulk masses are a linear combination of the identity matrix and the Yukawa times its Hermitian conjugate (or vice versa), 
which is the only spurion with appropriate quantum numbers. 
 The novel feature is that even if the Yukawa matrices are anarchical, 
the flavor hierarchy is obtained at low energies with protection from flavor violating processes due to partial alignment.

We now show that this same mechanism can be applied to the lepton sector to suppress flavor violation, despite the anarchic Yukawas and the low KK scale.
We will assume the bulk masses are linear combinations of the identity and appropriate combinations of Yukawas consistent with the assumed $SU(3)_L \times SU(3)_E \times SU(3)_N$ flavor symmetries, where the first group acts on lepton doublets, the second on charged lepton singlets, and the third on right-handed neutrinos.

We also assume a gauged lepton number symmetry. Since there are no massless gauge bosons that correspond to U(1)$_{\rm L}$ the lepton number symmetry
 has to be broken. Furthermore, to prevent overly large neutrino masses we must have the lepton gauge symmetry  preserved in the IR or broken only
 by a very small parameter (which would render a light gauge boson).
 This suggests that within our framework, probably the only consistent way to get small neutrino masses is when lepton number symmetry is broken
 on (or near) the UV brane so that lepton-number violation is sequestered from the IR.
 This argument rules out models with Dirac neutrino masses (see {\it e.g} \cite{RSfermions})
unless the RH neutrinos are localized towards the IR brane~\cite{DiracMajorana} if the symmetry breaking couples directly to the right-handed neutrinos.
 However, in such a case the LH doublets would have to be localized sharply near the UV brane
 to suppress their coupling to the Higgs and ensure that the resulting Dirac masses are small enough.
  This would also imply that the charged lepton masses are at least equally suppressed and is in conflict with data, given a full RS volume, since that tau
  mass is only two order of magnitude below the electroweak scale and $\tau_R$ cannot be too composite because of electroweak precision constraints. One possible alternative is to choose charges such the lepton breaking does not couple directly to the right-handed neutrinos.
We choose instead to focus on see-saw models with Majorana neutrinos in which lepton number symmetry is gauged in the bulk and broken 
on the UV inducing  RH Majorana neutrino masses~\cite{hsMaj}.


\section{The Model}
Lepton flavor can be explained via wavefunction overlaps only when 
  the Standard Model fields $L,E$ (weak doublet and charged singlet respectively) and  neutrinos $N$ reside in the bulk. We will assume this is the case and consider scenarios either with the Higgs field $H$ in the bulk or localized on the IR brane. We extend the mechanism suggested in~\cite{FPR} of anarchical QMFV 
to show how to construct a viable model of anarchic neutrino masses and mixing via a similar concept of
 lepton minimal flavor violation (LMFV). Under this assumption, the only flavor-violating structure is that of the Yukawa couplings. Bulk masses do not contain independent flavor violation and contain only Yukawa-dependent flavor violation determined   by their flavor transformation.
This means that
\be
C_E={\bf 1}+a k^2 Y_E^\dagger Y_E\,.
\ee
\be
C_N={\bf 1} + b k^2 Y_N^\dagger Y_N\,.
\ee
\be
C_L={\bf 1}+a_E k^2 Y_E Y_E^\dagger+b_N Y_N Y_N^\dagger\,.
\ee

LMFV leads to sufficiently suppressed lepton flavor violation, even with large Yukawa mixing entries, and also helps explain the relatively small hierarchy of neutrino masses. Furthermore we will see that with this assumption the difference between small mixing angles and nondegenerate wavefunctions vs degenerate wavefunctions and large mixing angles corresponds to a ratio of parameters in lepton vs. quark sector (that is $a$s and $b$s characterizing SU(3) symmetry-breaking) of about 0.1.

We assume neutrino masses arise from a see-saw mechanism involving heavy right-handed zero modes of $N$. Without a lepton number symmetry, masses would in general also arise from a higher-dimensional Majorana mass term involving $L$ and the Higgs field (but no right-handed neutrino) on the IR brane. However, 
since the bulk physics includes a quantum gravity sector we cannot impose a global lepton number symmetry in the bulk.
 Gauging such a symmetry without breaking it or breaking only at a low scale would imply a massless gauge field which is in conflict with constraints. We therefore assume lepton number symmetry broken only on the UV brane, permitting Majorana masses involving heavy right-handed neutrinos but sequestering lepton-number breaking from the IR. Notice that in the absence of tuning this implies a large Majorana mass for the right-handed neutrino and hence a see-saw mechanism. 

We assume that the bulk respects $U(3)_E\times U(3)_L\times U(3)_N$ symmetries 
(which includes the gauged lepton  symmetry), where  $E$ stands for charged lepton singlets, $L$ for lepton doublets, and $N$ for singlet neutrinos. Any flavor gauge anomalies can be cancelled by Chern-Simons terms or spectators on the UV  brane \cite{ArkaniHamed:2001is}.
In fact in the CFT dual theory the flavor gauge symmetries correspond to global symmetries and therefore it is not surprising
that the 5D anomalies can be straight forwardly cancelled.

The measured largeness of neutrino mixing angles seems to imply  comparable entries for all Yukawa matrix entries.  Mixing between the 1st and 3rd generations might in fact be measured to contradict this assumption, but for now the value is consistent with random large 
entries~\cite{anarchy} which we refer to as {\it anarchic}. This anarchical nature of the neutrino parameters suggests that to leading order the $U(3)_E$ flavor symmetry is  broken in the bulk  whereas
the $U(3)_L$  symmetry is not broken or  broken only by a small amount in the bulk and therefore broken primarily in the IR by the 5D Yukawa coupling. For this reason,  left-handed neutrino wavefunctions are fairly degenerate and don't induce small mixing angles as they do in the quark sector where small angles are correlated with mass ratios through left-handed wavefunctions.   We furthermore assume that the UV lepton breaking is flavor blind. 

As  in Ref. \cite{FPR}, which discussed quark flavor, we assume anarchic Yukawa matrices (that is, all entries of the same order) for both the charged leptons and the neutrinos. In the case of quarks, we found mixing angles that naturally aligned with masses since both arose from the quark wavefunctions. In this case we assume sufficient degeneracy among left-handed leptons that the mixing angles remain order unity as is the case in the five-dimensional Yukawa matrix we start off with.

The relevant gauge-invariant
5D Lagrangian  is:
\be
\CL &=& \CL_{\rm kin} +  \left(
k c_{L} \bar{L} L -k C_{N} \bar{N} N
 -k C_{E} \bar{E} E\right) \nn \\
  &&+\left( L H Y_{E} E + L \tilde{H} Y_{N} N\right)_{\rm IR}+\left(M_N N^T N\right)_{\rm UV}+h.c \,,
\ee
where capital $C_x,Y_{E,N}, M_N$ are general matrices and the small $c_i$ are universal.
Assuming for simplicity that any brane-localized kinetic terms are negligible, we can explicitly KK reduce.  We shall treat the UV and IR flavor and lepton number breaking perturbatively (we shall discuss this in more detail below).  
For simplicity we assume that the Higgs is localized on the IR brane although in general we expect that the resulting constraints
are more stringent in that case~\cite{abp} even though this depends on the details of the model (recently ~\cite{CFW} showed that in certain models with a bulk Higgs the constraints in the quark sector are comparable to models with an IR brane Higgs).

The charged lepton mass matrix is 
\begin{equation}
m_{E}\simeq2v k\, f_L  Y_{E} F_{E}\label{4Dmass}\,,
\end{equation} 
where $F_x$ correspond to the values of the lepton zero-modes
on the TeV brane. 
More explicitly,  the eigenvalues $f_{x^i}$ 
of the $F_x$ matrices are given by~\cite{aps}
 $f_{x^i}^2=
(1/2-c_{x^i})/( 1-\epsilon^{ 1-2c_{x^i} }
 )\,,$
where
$c_{x^i}$ are the eigenvalues of the $C_x$ matrices, $\epsilon=\exp[-\kpirc]$,
$\kpirc=\log[M_{\rm \bar Pl}/{\rm TeV}]$, $M_{\rm \bar Pl}$ is the reduced Planck mass
and $v\simeq 174{\rm GeV}$.
(we follow  here the convention 
where $0<f_i<1$ where according to their CFT interpretation $f_i$
corresponds to the amount of partial compositeness.
For KK states then we simply have the corresponding $f$ equal to unity. This is the reason for the factor of 2 in the Yukawa interactions)
The light neutrino mass matrix depends on the size of the wavefunctions of the RH neutrinos on the UV brane.
We shall denote them as $f^{UV}_N$ with  $(f^{UV}_N)^2=(1/2-c_{N})/( \epsilon^{ 2c_{N}-1}-1 
 )\,.$
\begin{equation}
m_{\ell}\simeq  m_{\nu}^T ({M^{\rm Maj}_{N}})^{-1} m_{\nu} \approx4 v^2 k^2  f_L \, Y_N^T F_N^T \left(2 F^{UV}_N M_{N} F^{UV}_N\right)^{-1} F_N Y_N f_L \,,
\end{equation} 
where to have a perturbative KK reduction we assume that $M^{N}\sim M_{\rm Pl}/10$ where $M_N$ is due to lepton breaking on the UV brane.
Since both $f^{UV}_N$  and $f_N$ are a function of the same $C$ we can solve this equation to find  $C$, assuming that $Y_{N}$ and $M_N$ are non-hierarchical, where if $C_N$ is universal this amount to solving only a single equation.

We see that the mixing angles are determined entirely by $f_L Y_N$. If $f_L$ is very nondegenerate, we would have mixing angles aligned with masses as with the quark case. However, if $f_L$ is very degenerate as we assume, mixing angles are determined  by the Yukawas and will be anarchic.

\subsection{Degenerate vs. Nondegenerate Right-Handed Neutrino Masses}

We have seen that large mixing angles requires fairly degenerate $c_L$s whereas charged lepton masses require some hierarchy in the $c_E$s. It is of interest to note that neutrino masses are fairly degenerate and furthermore depend on $Y_N^2/M_N$. The small hierarchy in the neutrino masses might be attributable to either nondegenerate right-handed neutrinos or to a small hierarchy in the Yukawa matrix that is nonetheless consistent with a fairly anarchical structure.

If the right-handed neutrinos are degenerate, it raises the interesting possibility that flavor-symmetry in the right-handed neutrino sector is violated only in the IR and only through the neutrino Yukawa.  The smallness of flavor-changing processes in the charged lepton sector seems to support such a possibility, since as we will see in the following sections, flavor-changing processes for charged leptons (with LMFV assumed) arise only through the $Y_N$s. Without the symmetric neutrino assumption, constraints from flavor violation (discussed below) typically  require $Y_N k$ of order 0.1. But an even more natural explanation for suppression of flavor-violation arising from $c_L$ at tree level would be the  suppression of all bulk $Y_N$-dependent terms, so that $c_N$s are degenerate and $c_L$ nondegeneracy arises only from $Y_E$.

Both cases are possibilities and we discuss flavor predictions in both cases. However for numerical simplicity we present an example only with degenerate right-handed neutrinos below.

\subsection{Example}

To give a concrete example we focus on the case where only the $U(3)_E$ flavor symmetry is broken in the bulk whereas flavor symmetry for the RH neutrino and LH leptons is broken only in the IR. That is, flavor for the latter two is broken by the Yukawas but not by the bulk masses. This is for convenience in the case of the bulk right-handed neutrino matrix but is necessary for order unity mixing angles in the case of the left-handed leptons. In the following subsection we determine the necessary degree of degeneracy of the  $c_L$s consistent with O(1) mixing.

We present an example for the normal hierarchy using  the best fit values from~\cite{Nureview}  
\be
m_{\nu^1}\leq 0.002 {\rm\, eV}, \,
m_{\nu^2}\simeq 0.009 {\rm \, eV}, \,
m_{\nu^3} \simeq 0.05 {\rm \, eV}\,.
\ee
where for concreteness we have set the lightest neutrino mass to its maximal value.
For the mixing angles we use the following values
\beq
\theta_{12,23,13}=35^o\,,\ 42^o\,, \ 8^o\,,
\eeq
where we have set $\theta_{13}$ to its maximal possible size at 1 $\sigma$ confidence level.
We are working in the convention where the MNS matrix is (setting the possible Majorana phases to zero for simplicity)
\begin{eqnarray} 
V_{\rm MNS}=
\bpm
1&0&0\cr
0&C_{23}&S_{23}\cr 
0 &-S_{23}&C_{23}
\epm 
\bpm
C_{13}&0&S_{13}e^{-i\delta_{CP}}\cr
0&1&0\cr 
-S_{13}e^{i\delta_{CP}}&0&C_{13}
\epm 
\bpm
C_{12}&S_{12}&0\cr
-S_{12}&C_{12}&0\cr 
0&0&1
\epm \,,
\end{eqnarray}
where $S_{ij}\equiv \sin\theta_{ij}\,,$  $C_{ij}\equiv \cos\theta_{ij}\,.$
In the following we give a concrete example for a viable model.
 
The value of the 5D Yukawa matrices are given by
\begin{eqnarray}
{\rm diag}( Y_{E}\,k) &\simeq& i(1.1,2.2,2.7)\,,\nonumber\\
{\rm diag} (Y_{N}\,k)  &\simeq&  -(0.02,0.03,0.07)\,.
\end{eqnarray}
and the Majorana mass on the UV brane is  $M^{UV}_N=M_{Pl}/10\,.$
We can express $c_E$ as a function of $Y_E$:
\be
C_E = 0.81 \times {\mathbf 1}_3-0.044\times|{\rm diag}(Y_E k)|^2\,, \ c_L=0.58 \,, f_L=0.016\,, c_N=0.27\,, f_N=0.48 \,,f_N^{UV}=0.00016\,.
\ee
Finally in the basis where $Y_E$ (and $C_E$) are diagonal then $Y_N$ is 
\be
Y_N  \to V_{\rm MNS} \, {\rm diag}(Y_N)\,.
\ee
These values were chosen to be consistent with the constraints from precision electroweak tests and also from lepton flavor violation that we discuss below.

\begin{table}[!hbt]\begin{center}
 \begin{tabular}{||c|c|c||}
    \hline\hline
    { Flavor}&   $c_E$ &$f_E$\cr
    \hline\hline
    I &   0.75&4.5\,10$^{-5}$\cr \hline
    II& 0.60&0.0094\cr\hline
    III&  0.50&0.15\cr
\hline\hline
 \end{tabular}
\caption{{\small The eigenvalues, of $C_x,F_x$ which yield the right masses and MNS elements at the 
 TeV scale~\cite{hs}. Note that $c_L,f_L=0.58,0.016$, $c_N,f_N,f_N^{UV}=0.27,0.48,0.00016$ and 
 $M^{UV}_N=M_{pl}/10$ are universal. Since in this example we assume that the $U(3)_{L,N}$ flavor group is only broken on the IR brane.}}
\end{center}\end{table}

\subsection{Large Mixing Angle Constraint on $c_L$s}

As we are going to see  bulk SU(3) symmetry which is exact or which is broken only by Yukawas (LMFV) leads to models which are
much safer from the point of view of CP and flavor violation observables than the generic case considered 
in Ref.\cite{abp}.
We require that the wavefunctions for the left-handed fields, and hence their bulk masses, are sufficiently degenerate to reproduce the neutrino anarchical mixing without tuning.

We now calculate how degenerate the bulk mass parameters should be. Suppose that the bulk theory is not exactly SU(3)$_L$ symmetric.
Let us estimate   the maximal allowed deviation from that limit such that 
the MNS matrix is still anarchic.
The MNS mixing angles are controlled by the ratio of the $f_L^{i}s$.
For concreteness let us demand that the corresponding ratio between the  $f_L^{i}s$ never deviates from 1 by more  than 10\% so that the angles are related to the 5d Yukawa matrices and 
not the ratio of wavefunctions as it would be when they are very nondegenerate as they were in the quark case considered in 
Ref.~\cite{FPR}.
 We have already pointed out that in order to avoid tension with precision electroweak constraints we need  left-handed and charged lepton wavefunctions localized in the UV. 
One can compute the corresponding deviation from degeneracy allowed in the $c_L^{i}\,.$
We find that $\delta c_L/c_L \sim 0.8\%\,.$
This constrains the flavor-breaking contribution to the $c_L$s from the Yukawas to be of this size.

The rough size of this restriction is easy to understand, as it is comparable to  the inverse of the volume of the space. The point is once the $c_L$s are degenerate within this amount, the strong exponential dependence of the wavefunctions disappears and the wavefunctions are more degenerate.
We note that this restriction on the Yukawa contribution to the left-handed bulk mass suppresses flavor violation, since the 
$Y_N$-dependent terms in $C_L$ induce flavor-changing processes, this is further explored below. 

\subsection{Electroweak Constraints}

In general the size of the IR wavefunctions is constrained by precision electroweak measurements since observed branching fractions of the $Z$ agree very well with Standard Model predictions. By choosing $c>1/2$ the wavefunctions on the IR brane are guaranteed to be small and the constraints determine how small they need to be. Since the measurement of $Z\to \mu \bar{\mu}$ is slightly better than the corresponding branching fraction measurement for $\tau$s, the constraint on the left-handed wavefunctions (which we are assuming are essentially degenerate) is slightly stronger than that on the wavefunction for the right-handed $\tau$. 

 Non universal coupling of the $Z$ to the fermions is highly constrained by LEP and other experiments~\cite{PDG}
\be
R_\mu = 20.785 \pm 0.033\,, \ \ \ \  
R_\tau =20.764 \pm 0.045\,, 
\ee
note that the above constraints related to $Z\to \tau\tau,\mu\mu$ are roughly 50\%,100\% more stringent than the corresponding constraint from $Z\to b\bar b$, $R_b = 0.21629 \pm 0.00066$, respectively.
In our model the hierarchy in the charged lepton masses would be explained by localization of the charged lepton singlets
in different points in the bulk. 
This implies that the coupling to KK $Z$
is generation-dependent, which in turn results in a 
non-universal shift in its coupling to the SM $Z$ via mixing
of KK $Z$ with zero-mode $Z$:
$
\delta g_Z^{l^i} \sim g_Z^{l^i} f^2_{l^i} 
\kpirc \frac{ m_Z^2 }{ m_{\rm KK }^2 }$
where $\kpirc=\log \left( M_{ Pl } / \hbox{ TeV } \right) $, $l^i$ stands for a LH or RH  lepton of $i$th generation.
The constraint from the $Z$ branching fractions to $\mu$ and $\tau$ imply that the maximal value allowed for $,f_L,f_{E^3}$ are   
\be
f_L\lsim 0.13\,, \ \ \  f_{E^3}\lsim0.15 \,.\label{range1}
\ee
Setting both to their maximal values we find that to get the right tau mass we require
\be
(\bar Y_E)_{33} \equiv 2 k \,{\rm diag}(Y_E)_{33} \gsim 0.5\,.\label{YEbound}
\ee
Saying it differently in the case where we set the average value of $\bar Y_E$ to its maximal perturbative value (allowing for at least three KK modes in the effective theory), $\bar Y_E\sim4$~\cite{abp,aps},
we find the correlated ranges for $f_L$ and $f_E^3$:
\be
0.016 <f_L< 0.13\,, \ \ \  0.019< f_{E^3}<0.15 \,.\label{range2}
\ee
where we have put together the  two precision constraints  with the $\tau$ mass to determine the allowed range of left-handed wavefunctions and right-handed charged lepton wavefunctions.  

%

In cases where the gauge symmetry is enhanced to ensure that
the model passes the constraints from  oblique parameters~\cite{cust1} and $Z\to b\bar b$~\cite{custZbbar}
other contributions are expected.
These would be
from the exchange of KK modes of the
extra $U(1)$ arising from the extended $5D$ gauge symmetry.
A second type of correction to $Z\to l\bar l$ is due
to $\ord{1}$ mixing between the lepton zero modes and the KK modes.
 This contribution is of
order $ (2 k v  Y_{N}  f_{l^i}/m_{KK})^2$. 
We will see that constraints from lepton flavor violation require rather small values for $ Y_{N}$   such that the contributions of the second type are subdominant.


We also need to consider potential constraints from the invisible $Z$ decay width due to mixing of the active neutrino
with sterile states.
To see why the light neutrino mixes with the KK sterile ones
let us focus on the corresponding $3\times3$ mass matrix for the third generation only treating the UV Majorana
term as a perturbation where for simplicity we discuss  the mixing within a single generation,
\begin{eqnarray} \bpm
L &N& N^{(1)} & {N^{c}}^{(1)}\epm 
\bpm
0& \bar Y_N v f_L f_N& \bar Y_N v f_L&0\cr
\bar Y_N v f_L f_N & M^{\rm Maj}_{N^{(0)}}&M^{\rm Maj}_{N^{(01)}} &0\cr
\bar  Y_N v f_L &\left(M^{\rm Maj}_{N^{(01)}}\right)^T & M^{\rm Maj}_{N^{(1)}} &M_{KK}\cr 
0 &0&M_{KK}&0
\epm 
\bpm
L \cr N\cr  N^{(1)}\cr {N^{c}}^{(1)} 
\epm
\,,
\end{eqnarray}
where $M^{\rm Maj}_{N^{0}}= 2(F_N^{UV})^T M_N F_N^{UV} \sim 3 \times 10^{12}$GeV, $M^{\rm Maj}_{N^{(1)}}= 2(F_{N^{(1)}}^{UV})^T M_N F_{N^{(1)}}^{UV}$,  $M^{\rm Maj}_{N^{(01)}}= 2(F_{N^{(1)}}^{UV})^T M_N F_N^{UV}$,  $F_{N^{(1)}}^{UV}$ is the UV value for the WF of the first KK RH neutrino
 and $m_D\equiv \bar Y_N v f_L f_N\sim 1\,$GeV and also the 31 entry is $m_D^{(1)}\equiv \bar  Y_N v f_L$ is of a similar size.
To analyze the constraints on our model we describe how to approximately diagonalize the
neutrino mass matrix. We first focus on the block that contains the largest entries related to the RH zero modes and excited modes which is induced via the UV term, neglecting EWSB.
To bring it to a diagonal form a 2x2 rotation matrix is required where the mixing angles, $\theta_N$, is sizable of ${\cal O}(F_{N^{(1)}}^{UV}/F_N^{UV})\sim 0.5$.
After this rotation applied the mass matrix is:
\begin{eqnarray}
\bpm
0& \bar  Y_N v f_L( c_N f_N - s_N)&\bar  Y_N v f_L(s_N f_N + c_N)&0\cr
\bar  Y_N v f_L( c_N f_N - s_N) &  M^{\rm Maj}_N&0 &s_N M_{KK} \cr
\bar  Y_N v f_L(s_N f_N + c_N)&0 &0&c_N M_{KK}\cr 
0 &s_N M_{KK}&c_N M_{KK}&0\epm
\,,
\end{eqnarray}
where $s_N,\,c_N \equiv \sin \theta_N,\, \cos \theta_N\,.$
The 3x3 block related to $N,N^1$ and ${N^{c}}^{(1)}$ now has one big eigenvalue and two of a TeV (KK mass) size.

We see from this matrix that we have a see-saw mechanism between $L$ and the heavy eigenstate (the linear combination of $N$ 
and $ {N^{c}}^{(1)}$) and furthermore that the light weakly charged state (that is mostly $L$) mixes with a linear combination of the singlets $N$ and $ {N^{c}}^{(1)}$.
 
Defining $\bar m_D\equiv \bar Y_N v f_L( c_N f_N - s_N)$ and  $\tilde m_D\equiv\bar Y_N v f_L(s_N f_N + c_N)$ we find a light eigenstate of mass $\bar m_D^2/{M}_N^{Maj}$ and a mixing of the light eigenstate with the singlet eigenstate  
of order ${\cal O}(\tilde m_D/M_{KK})$ which must be less than 1\%.  
This value is constrained by precision $Z$ invisible width measurements 
$n_{\rm eff} = 3-\sum_n \sin^2\theta_n$ with $\delta_n\lsim 5\times 10^{-3}$. In our model the sum over the three generation yields 
$\delta_n\lsim 7\times10^{-4}$ over the whole range
given in Eqs. (\ref{range1},\ref{range2}), so we are already below the exprerimental bound.
In our concrete example and numerical calculations we have, for simplicity, set $s_N$ to zero.

\section{Lepton flavor violation}

Having considered precision electroweak constraints, we now turn to potential lepton flavor violation in our model. It should be borne in mind that we are in a situation that superficially appears extremely dangerous. Our model contains anarchic Yukawas as well as KK modes as light as a few TeV. One would naively expect large flavor-violating effects.  In fact Ref. \cite{abp} demonstrated that without LMFV and our SU(3) flavor assumptions that the KK scale would need to be raised to roughly 7 TeV with a bulk higgs~\footnote{The bound quoted in~\cite{abp}, 15.8 TeV, is obtained by adding coherently five independent contributions. If we switch on each of these separately the resulting bound is reduced by
a factor of $\sqrt{5}$, we thank Kaustubh Agashe for pointing this to us}.
An even worse bound applies with an IR brane localized Higgs, in which case the contribution to $\mu \to e\gamma$
is UV sensitive and   without flavor symmetries would require a cutoff scale as high as 80 TeV.

We now demonstrate that our assumed LMFV and a slightly smaller value for the neutrino Yukawa than the maximal allowed value, in combination with the relatively small wavefunctions for the charged leptons, implies that lepton flavor violation is sufficiently suppressed.   We will also consider future directions for flavor-changing processes and  CP-violating observables.

The most severe constraints involve charged lepton flavor violating processes.
Contributions to charged lepton flavor violation arise due to misalignment between the couplings to the physical $Z,W$ and photon and the masses and arise
through mixing with KK fermion and gauge bosons.
Without LMFV, the most severe constraint would be due to the mixing of the LH charge leptons   with the KK RH charge leptons.
This mixing is proportional to the Higgs Yukawa coupling on the IR brane.
 Recall that the mass matrix square is proportional to $Y_E F_E F_E^\dagger Y_E^\dagger$.
On the other hand the above mixing is controlled only by the Yukawa, $Y_E Y_E^\dagger$.
In the general case we do not expect $Y_E$ and $F_E$ (which is a function of $C_E$, the bulk masses)
to commute. Hence the two contributions  will in general be  misaligned and far too large.
  However, this contribution is highly suppressed with our LMFV assumption, since $Y_E$ and $C_E$ are simultaneously diagonalized.

Let us explore this in slightly more detail.
A suppression for this contribution could arise be due to hierarchies in the values of the LH fields on the IR, $f_L$. 
Such a suppression is absent in our case since the value of wave functions of the LH leptons as well as
the eigenvalues of the Yukawa matrices are of the same order, due to the anarchical nature of the neutrino parameters.
Thus, the resulting $Z$  flavor violating coupling, say $\mu\to e$ which is most severely constrained,
is of order
\be
\left(g_Z\right)_{12} \approx  \left(2 f_L  k Y_E v\over M_{KK}\right)^2_{ij} (U_L)_{1i} (U_L)^*_{2j} \,.
\ee
where $U_L$ is the rotation matrix between the basis in which the bulk masses are diagonal and the basis
in which the Yukawa matrices are diagonalized (from the left).
Without LMFV, this would be too large. But  with LMFV, which ensures that  $c_E$ (and hence $f_E$) is a function only of $Y_E$,
the above contribution vanishes since the both the Yukawa couplings and bulk masses can be simultaneously diagonalized.

There can also be a tree-level flavor-changing contribution when $c_L$ is not entirely proportional to the identity but has a piece consistent with LMFV proportional to $Y_N^\dagger Y_N$. This term is suppressed in order to get large mixing angles, but in principle might still be present.    Recall that due to LMFV we expect 
$c_L={\mathbf 1}+ a_E\,k^2 Y_E Y_E^\dagger + b_N\, k^2 Y_N Y_N^\dagger\,. $  
Clearly, only a term proportional to $Y_N Y_N^\dagger$  induces flavor violation.

This term could contributes to flavor-violation at tree level through the exchange of  a KK $Z$. Relative to the loop level effects we consider below, the constraint is roughly $16 \pi^2 (2 \pi k r_c)$
 times as strong (both share the suppression factors of  $(m_Z/m_{KK})^2 f_L^2$). Even with a constraint on $b_N$ of order $0.01$ as we found from considering large mixing angles, this would be too big unless $Y_N k$ itself is small. 
 We will see below that   loop effects constrain $Y_N k$ to be order order 0.1, which would be acceptable.  

Note that this tree-level constraint from $C_L$ is relevant only when neutrino-flavor symmetry is broken in the bulk. In the more symmetric case in which neutrino-flavor-symmetry is broken only in the IR, the tree-level constraint just considered is automatically satisfied. There is no tree-level constraint on $a_E$ from flavor-changing effects since $Y_E$ and the charged leptons masses are aligned.

\subsection{$\mu\to e\gamma$}
\begin{figure}[htbp] 
   \centering
   \includegraphics[width=1.7in,angle=270]{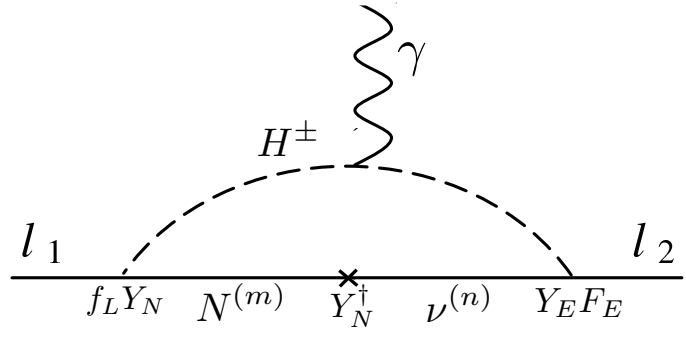} 
   \caption{Contributions to $\mu\to e\gamma$ via mixing on the IR brane.}
   \label{diagram}
\end{figure}

Because in the absence of $Y_N$, the charged lepton mass matrix and interactions can be simultaneously diagonalized, the leading contributions to charged lepton flavor-changing processes   must involve $Y_N$. The process $\mu \to e \gamma$  can   occur only at loop level.
The most dangerous contributions are due to $\mu\to e \gamma$~\cite{abp}. The relevant one loop diagram is shown in figure 
\ref{diagram}. \footnote{Note that this diagram is different from the one calculated in \cite{abp}
since in our case LFV is due only to the neutrino Yukawa matrix so the photon has to be attached to the charged Higgs internal line.}
The contribution is conventionally written as
\begin{equation}
A(\mu\rightarrow
e\gamma)=\frac{-iC_{L,R}\,}{\,2m_\mu}\,\overline{u_{L,R}}(p^{'})\sigma^{\mu\nu}u(p)\,F_{\mu\nu}\,, \label{ampmueg}
\end{equation}
Following~\cite{abp,Ng} we can write the current bound on the BR as~\cite{PDG}
\be
BR(\mu\rightarrow e\gamma)=\frac{12\pi^2}{(G_F m_\mu^2)^2}\sum_{i=L,R}
|C_i|^2\leq 1.2 \times 10^{-11},
\label{mueg}
\ee
As pointed out in~\cite{aps,abp} with a brane higgs the contributions are logarithmically divergent and in term of spurions proportional to 
\be
C_L \propto 8v k^3  f_L Y_N Y_N^\dagger Y_E F_E = 4k^2\, m_\mu \left(Y_N Y_N^\dagger\right)_{12} \,,
\ee
where in the RHS of the equality we have included only the dominant contribution  proportional to the muon mass
and $\left(Y_N Y_N^\dagger\right)_{12} $ is evaluated in the charged lepton mass basis, where $Y_E$ is diagonal. Note that in our model we expect the operator with RH chirality to be suppressed by $\ord{m_e/m_\mu}$ according to the assumption that only right-handed fields account for masses.
Due to the fact that the contribution is diverging the precise value of $C_i$ is not physical
since it would be matched to a counter term in the form of a higher dimensional operator.
We compute the amplitude in order to establish that it is 
log divergent and also since it would be useful  for the case of a bulk Higgs where the contribution is finite.
For simplicity we assume that the RH and LH KK masses are of same size. Furthermore in the limit of large $N_{KK}$ (or equivalently large $(\Lambda/k)$, the log term would be the dominant contribution. When $N_{KK}$ is not large, the loop log should nonetheless give a reasonable estimate of the net contribution, as we discuss further in the following section.
\begin{eqnarray}
A(\mu\rightarrow e\gamma)&=& B  \sum_{i,j}\int \frac{d^4k}{(2\pi)^4}  (2k^\mu -q^\mu) A_\mu \overline{u_L}(p^{'})
\left[i\frac{\slsh{k}-\slsh{p}+M_{KK}^{(i)}}{(k-p)^{2}-M_{KK}^{(j),2}}\right]^2 u(p)
\cdot\frac{i}{k^2-m_W^2} \frac{i}{(k-q)^2-m_W^2} 
 \,,
 \end{eqnarray}
where $B=8  e m_\mu  v k^3\,  f_L Y_N Y_N^\dagger Y_E F_E, \ q={p}^{'}-p$ and we have used the mass insertion approximation for the internal neutrino KK line.  
After integrating over the 4D momenta, the leading order contribution (setting subleading external momenta to zero) is 
\be\hspace*{-.95cm}
A(\mu\rightarrow e\gamma)&=& {B\over 2}\, \overline{u_L}(p^{'})\,\sigma^{\mu\nu}F_{\mu\nu}u(p) 
 \sum_{i,j}\int_0^1 dx_1 \int_0^{1-x_1} dx_2 \int_0^{1-x_1-x_2}  { dx_3\,(-1 + x_1 + x_2 + x_3)\over 48 \pi^2 (m_W^2 (-1 + x1 + x2) - M_{KK}^{i,j,2} (x1 + x2))}\nonumber\\
 &=& \sum_{i,j}{ e m_\mu^2  \over  (24 \pi)^2 (M_{KK}^{i,2}+M_{KK}^{j,2})} \,( \bar Y_N \bar Y_N^\dagger)_{12} \,\overline{u_L}(p^{'})\sigma^{\mu\nu}F_{\mu\nu}u(p ) +\ord{m_W^2/M_{KK}^2}
\,,
\ee
where  $\bar Y_N\equiv 2k\, (Y_x)\,.$ 
For simplicity we had set the two KK masses to be equal. Notice that, since $M_{KK}^{n,2}\sim \left(n M^{(1)}_{KK}\right)^2$, the double independent summation of the two
KK masses induces a logarithmic divergence $\sim \log N_{KK}$ as was anticipated by \cite{abp}.
Interestingly, unlike the diagram considered in \cite{abp} which vanished for zero Higgs mass, the loop diagram we consider here does not vanish in the limit $m_{H,W}\to0$
which was taken in the last step.  Generically both diagrams are present but in our model due to LMFV only
the above contribution induces lepton flavor violation.
As anticipated, the contribution is UV sensitive and we can estimate its size by simply considering the higher dimensional
operator on the TeV brane.

\subsubsection{UV Sensitivity With and Without Flavor Symmetry }

Let us consider the forms of the relevant higher-dimensional operators that are necessary for matching to the 5D theory in order  to understand the forms and magnitudes of the expected counterterms (for similar considerations see also the analysis of \cite{AgasheNDA}). 
The leading operator is of the form $O^{IR}_{\rm dipole}\propto  H F^{\mu\nu} \bar L \sigma_{\mu \nu} E\,.$
Since we are interested in an electromagnetic dipole operator it should be proportional to  $g_5^e$.
Furthermore, since the operator requires a Higgs interaction between a SM SU(2) doublet and singlet field 
it is proportional to $Y_E$ (the spurion that breaks the corresponding flavor symmetry)\footnote{When we discuss the generic case
the flavor symmetries are assumed to be  badly broken. This amounts to setting the 5D Yukawa to at or near its largest possible value. This is done below demanding
at least three KK state before the Yukawa couplings hits a Landau pole}.

We first note that the Yukawa and gauge couplings are dimensionful parameters, indicative of the nonrenormalizable nature of the five-dimensional theory. The 5D Yukawa interactions are of the form $H Y_E \bar L E$, say for the charged leptons.
Note that since $E$ and $L$ are bulk fields $Y_E$ is of mass dimension $-1$. 
The 5D gauge interaction are given by  $g_5 \bar E \gamma_M E A^{M}$ ($M=0,…,4$)
where in the case since all the particles involved in the interaction are bulk fields $g_5$ have mass dimension $-1/2$. 
Since the couplings are dimensionful, to ascertain the degree of divergence, it is not enough to consider only the dimension of the fields in an operator, but we must also consider operators with different numbers of insertions of couplings independently since they have differing degrees of divergence.

Let us first consider the operator with a brane-localized Higgs field, which takes the form
\begin{equation}
g_5 \left({1 \over \Lambda_{IR}^2}\right)_{ik}   (Y_E)_{kj}\, H F^{\mu \nu} L^i \sigma_{\mu \nu} E^j\,,
\end{equation}
where the brane-localized Higgs has dimension $1$, the 5d gauge coupling has dimension $-1/2$, the Yukawa has dimension $-1$, and we have treated the cut-off as a flavor-dependent matrix according to where each Yukawa gets strongly coupled so that $i,j,k$ are flavor indices.
LMFV tells us that these $\Lambda s$ must be cutoffs for the $Y_N$s or else the flavor structure could be rotated away.  Naively it would seem as if the bounds with universal doublet bulk masses and anarchical neutrino 5D Yukawa matrices, $Y_N$, would be quite severe. (Note that there is no  extra suppression from the ratio of wave functions on the IR of $\ord{m_e/m_\mu}$ 
that would apply in the generic case without LMFV that we will briefly review below)
 However, recall that in our case  the gauge flavor symmetries
in the limit where $Y_N\to0$ eliminate the flavor conversion altogether, so suppression by small $Y_N$ will suffice.

We determine $\Lambda_{IR}$ as follows:
We require that the one loop correction to the Higgs-KK-fermions
Yukawa couplings do not exceed the tree level one (see also {\it e.g.} \cite{APR,AgasheNDA,LRS}),
 which implies that
\be
(N^Y_{KK})^2   \bar   Y_x^2/16 \pi^2\lsim1\,, 
\ee
 where $N^Y_{KK} $
 is  the number of KK modes that is allowed before the Yukawa becomes strongly coupled, and as before, $\bar Y_x\equiv 2k\, (Y_x)$ where $x=E,N$.
 \footnote{Note that $N^Y_{KK}$ is not necessarily a physical parameter as some other coupling could in principle become strongly coupled at lower energy, thereby providing the true cut-off for the low-energy theory.} This corresponds to a 
 cutoff of the effective theory, $\Lambda_Y$ 
 \be
 \Lambda_Y\sim N^Y_{KK} \, k\lsim 4\pi k /  \bar Y_x \,,\label{LambdaY}
 \ee
 where note that here and below, we do not distinguish between the KK mass scale and $k$ (the former is slightly bigger). The precise relation clearly depends on the regularization 
 scheme~\cite{AgasheNDA} so we  simply take $M_{KK}\sim k$.
Substituting into our above estimate, we find an operator of the form
\be
O^{IR}_{\rm dipole}= g_5^e\,  \left(\bar Y_N  \bar Y_N^\dagger Y_E\right)_{ij} H F^{\mu\nu} \bar L^i \sigma_{\mu \nu} E^j/16 \pi^2 k^2 \,+\,h.c\,,
\label{5DO}
\ee
where the above fields aside from $H$ are 5D fields, and for $\mu\to e\gamma$ we have $i,j=1,2$ or $i,j=2,1$.
We can convert to the normalized four-dimensional fields ~(see {\it e.g}~\cite{RSfermions}) 
use the relation between the 5D and 4D gauge coupling (see {\it e.g}~\cite{APR}), using the mass relation given in (\ref{4Dmass}) 
$m_\mu \simeq \bar Y_E k v  f_{L^2}f_{E^2}$ to rewrite the above as
\be
 O^{IR}_{\rm dipole}\approx    {e \,m_\mu  \over2\,16 \pi^2 k^2} \,  (\bar Y_N \bar Y_N^\dagger)_{12}\,   F^{\mu\nu}\, \overline{ e_L} \,\sigma_{\mu \nu} \mu \,.
\ee
This operator has the same parametric dependence as the loop diagram. Notice that once we account for flavor symmetry, it is clear there are two insertions of the neutrino Yukawa. Since these are dimensionful, the operator with Yukawas inserted can depend at most logarithmically on the cutoff, which is what we found explicitly above.

Via Eqs. (\ref{ampmueg}, \ref{mueg}) we find the following value for the $BR(\mu\to e\gamma)$
\be
BR(\mu\to e\gamma)^{IR}\sim 
  \,4\times 10^{-8}\times \left(\bar Y_N \bar Y_N^\dagger\right)_{12}^2 \times \left({3\,{\rm TeV}\over M_{KK}}\right)^4\,.
\ee
Assuming order one mixing angles this relation is translated to the following upper bound on the value of the neutrino Yukawa coupling
\be
  (\bar Y_N \bar Y_N^\dagger)_{12}  \lsim \,0.02\times \left({3\,{\rm TeV}\over M_{KK}}\right)^2\,.\label{UVbound}
\ee
Since by assumption all the entries of $Y_N$ are of the same order, we interpret this bound as  $Y_N k$ is less than or of order 0.1 for the most natural models.
Note that the NDA estimate is bigger than the explicit loop contribution (which is also enhanced by a log), so it is most likely a fairly conservative estimate.

Our result differ from the more generic case (without LMFV) considered in \cite{abp} (for a detailed description of the structure see {\it e.g}~\cite{aps,abp}) as follows.
 As before
 the IR-brane, higher dimension operator is given by
\be
O^{IR}_{\rm dipole}= g_5^e\, \bar Y_E H F^{\mu\nu} \bar L^i \sigma_{\mu \nu} E^j/\Lambda_{IR}^2\,+\,h.c\,,
\label{5DO}
\ee
We   set $\Lambda_{IR}$ to the smallest of the cutoff  values obtained in the via the perturbative limits on 
the various couplings, given in Eq. (\ref{LambdaY}).
As before the above expression is simplified when when we switch from 5D fields to canonically normalized 4D ones.
We find that dipole cutoff operator can be rewritten (in terms of the zero modes and 4D couplings) on the IR brane as
\be
 O^{IR}_{\rm dipole}\approx   {e\,m_\mu \over2\Lambda_{IR}^2}\, max\left ({f_{L^1} \over  f_{L^2}} ,{f_{E^1}\over f_{E^2}}\right)  F^{\mu\nu} \bar \mu \,\sigma_{\mu \nu} e\,,
\ee
where we have replaced the Higgs by its vev. It is clear that the case where ${f_{L^i}}={f_{E^i}}$ would minimize the RS contributions that the above expression is reduced to
\be
 O^{IR}_{\rm dipole}\approx   {e\, \sqrt{m_\mu m_e} \over2\Lambda_{IR}^2}\,   F^{\mu\nu} \bar \mu \,\sigma_{\mu \nu} e\,,\label{4DO}
\ee
This is in agreement with the more generic case considered in Ref.~\cite{abp} and their rough estimate.

We now return to the LMFV scenario with a bulk Higgs, in which case the contribution to the dipole operator is actually finite and  UV sensitivity~\cite{aps,abp} appears only at higher-loop order as there is only one KK sum in the one-loop diagram \footnote{One can verify that with a linear Higgs profile, say as in the case of gauge Higgs unification then the 
overlap integral between to KK states of a different level is negligible.} and 
thus the resulting bound is weaker.
The loop contribution is the same as the one discussed above with a brane Higgs, with a single KK sum and therefore it is finite.
 We thus find the following bound on the RH Yukawa coupling
\be
  (\bar Y_N \bar Y_N^\dagger)_{12}  \lsim \,0.3\times \left({3\,{\rm TeV}\over M_{KK}}\right)^2\,,\label{1loopbound}
\ee
which is indeed weaker than the IR Higgs case found in Eq. (\ref{UVbound}).

 The higher-dimensional operator from the five-dimensional theory is smaller with a bulk Higgs because the Yukawa coupling has dimension $-1/2$. We now have
 $Y_N^2 k/16 \pi^2 \sim 1$ for strong coupling while we still have $\Lambda \sim N k$.
Therefore we can now write the higher dimension operator as
\begin{equation}
g_5 {1 \over \Lambda_{IR}}\, \left({1 \over \Lambda_{IR}}\right)_{ik}   (Y_E)_{kj}\,HF^{\mu \nu} L^i \sigma_{\mu \nu} E^j \sim g_5 (Y_N  Y_N^\dagger Y_E)_{ij} {1\over 16 \pi^2 \Lambda_{IR}} H F^{\mu \nu} L^i \sigma_{\mu \nu} E^j \,.
\end{equation}
Clearly this is smaller than any counterterm for a divergent contribution at this order.
We also see that at higher order, we would have divergent contributions (replacing the final $\Lambda_{IR}$ by ${Y^2 /16 \pi^2}$) but these are suppressed by a loop factor and yield a weaker bound.

\subsection{Subleading Constraints and Predictions}

Typically once the constraint from $\mu\to e\gamma$
is satisfied, the rest of the constraints are also satisfied since they are less severe. 
We will present bounds on $Y_N$ with the understanding that they tell us how much the experimental measurements need to be improve
to be competitive.
The next most stringent bound is from $\mu\to e$ conversion~\cite{abp} but we also consider
 $l^i\to \bar l^k l^k l^m$-type processes.
We parameterize these effects in the following 
effective Lagrangian~(we follow the notation of~\cite{abp,Ng,Kuno:1999jp}):
\begin{eqnarray}
-\mathcal{L_{{\rm eff}}}&=&
\frac{4G_F}{\sqrt{2}}\left[
g_4^{\mu e}(\bar{e}_L\gamma^\mu\mu_L)(\bar{e}_L\gamma_\mu e_L)+g_6^{\mu e}(\bar{e}_L\gamma^\mu\mu_L)(\bar{e}_R\gamma_\mu e_R)+{1\over2} g_{L}^{\mu e} (\bar{e}_L\gamma^\mu\mu_L) J_\mu\right] + {\rm h.c.} \,,
\label{Leff}
\end{eqnarray}
where  we have neglected the RH  flavor violating currents, which are flavor-conserving due to  LMFV, 
$G_F$ is the Fermi constant,
$J_\mu$ is the current relevant for $\mu\to e$ conversion.

 The strongest constraint on $\mu-e$  conversion is obtained
with $^{48}_{22}$Ti atoms,
The rate is~\cite{Kuno:1999jp,Ng,PDG}
\begin{equation}
B_{conv} = \frac{p_e E_e G_F^2 m_{\mu}^3 \alpha^3 Z_{eff}^4 [v^u (2Z+N)+v^d(2N+Z)]^2 F_p^2}{\pi^2 Z \Gamma_{capt}} \,
|g_{L}^{\mu e}|^2 < 6.1 \times 10^{-13}\,,
\label{ueconv}
\end{equation}
where $\alpha$ is the QED coupling strength,
$\Gamma_{capt}\sim 10^{-8}{\rm \,eV},\,
  p_e \sim E_e \sim m_\mu,\, Z_{eff} \approx 17.6,\, Z = 22,\, F_p \sim
0.55,\, N=26\,.$
where $v^{u,d}=\pm1/2-(4/3,-2/3) \sin^2 \theta_W$ and $ \sin^2 \theta_W\sim 0.22\,.$
The bound on $\mu\to 3e$ is~\cite{PDG}: 
\begin{eqnarray}
BR(\mu \rightarrow 3e) &=& 2 |g_{4}^{\mu e}|^2+|g_{6}^{\mu e}|^2 < 10^{-12} \,.
\label{bfracs}
\end{eqnarray}

\begin{figure}[htbp] 
   \centering
   \includegraphics[width=2.in,angle=270]{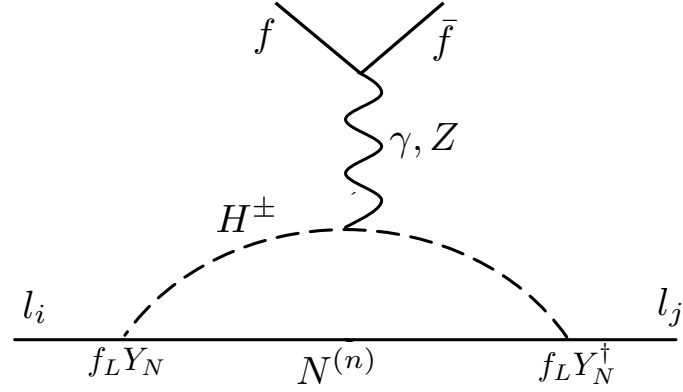} 
   \caption{Contributions to lepton flavor conversion.}
   \label{diagram2}
\end{figure}

With LMFV and true $c_L$ universailty  the dominant contributions to flavor-changing LH currents arise at loop level (see discussion on possible tree level contributions below).
In fig. \ref{diagram2} we show the diagram that induces the leading contribution to the above processes, it does not involved helicity flipping and the KK sum is finite (unlike for $\mu\to e\gamma$).
 The amplitude for the above processes is readily computed from the diagram (with external momenta in the loop set to zero)
\begin{eqnarray}
A(\mu\rightarrow e)&\simeq&  g_Z f_L^2 (\bar Y_N \bar Y_N^\dagger)_{12}   \sum_{i}\int \frac{d^4k}{(2\pi)^4}\, J_\mu 2 \overline{u_L}(p^{'})
\left[i\frac{k^\mu   \slsh{k} }{k^{2}-M_{KK}^{(i),2}}\right]  u_L(p)
\cdot\frac{i^2}{(k^2-m_W^2)^2}  \,,
 \end{eqnarray}
where $J_\mu$ corresponds to the two fermion current attached to the $Z$ propagator and $g_Z=g_2/2\cos \theta_W$ is the $Z$ coupling.
Subtracting  the log divergent contributions (which corresponds to the renormalization of the gauge coupling) and keeping the lowest KK contributions that dominate the sum we find
\begin{eqnarray}
A(\mu\rightarrow e)&\simeq&   g_Z f_L^2 (\bar Y_N \bar Y_N^\dagger)_{12} \, \left( {M_W\over 4 \pi M_{KK}}\right)^2 \left[1 + 2 \ln \left( {M_W^2\over M_{KK}^2}\right) \right] \, J_\mu\,  \overline{u}(p)\gamma^\mu u_L(p')\,.
  \end{eqnarray}
We then match the above result to the leading
contributions to the four fermion operators in Eq. (\ref{Leff})
\be
g_{L}^{\mu e}&\approx&  f_L^2  (\bar Y_N \bar Y_N^\dagger)_{12} \, \left( {M_W\over 4 \pi M_{KK}}\right)^2 \left[1 + 2 \ln \left( {M_W^2\over M_{KK}^2}\right) \right] \nonumber\\
g_{4,6}^{\mu e}&\approx& v^e_{L,R} \,g_{L}^{\mu e} \,,  \label{coeff} 
\ee
where $v^e_{L,R}=(-1/2,0)+ \sin^2 \theta_W\,$.
Thus  
\be
\label{LMFVco}
B^{LMFV}_{conv}\approx \left(9\times 10^{-20}\,,\,1\times 10^{-13} \right)\left[{\left(\bar Y_N \bar Y_N^\dagger \right)_{12}\over 0.02,0.3}\right]^2 \left(f_L\over 0.13,0.016\right)^4  \left({3\,{\rm TeV}\over M_{KK}}\right)^4\,,
\ee
where the numerical value is shown for the two extreme values of $f_L$ (see Eq. (\ref{range2})) and the upper bound for the 
neutrino Yukawa coupling from $BR(\mu\to e\gamma)$ for the Higgs on the brane and Higgs in the bulk, see Eqs. (\ref{UVbound}, \ref{1loopbound}). 
This can be translated to the following bound on the neutrino Yukawa coupling:
\be
\left(\bar  Y_N \bar Y_N^\dagger \right)_{12}\lsim(0.7,5\times 10)\times \left({3\,{\rm TeV}\over M_{KK}}\right)^2 \times \left({f_L\over 0.13,0.016}\right)^2\,,
\label{YNconv}
\ee
significantly weaker than the bounds obtained via $\mu\to e\gamma\,,$ in particular when $f_L$ is set to its minimal
value the size of $Y_N$ exceeds its maximal perturbative allowed value.
Similarly the predicted BR for $\mu\to 3e$ in our case is
\be
BR(\mu\to 3e)^{LMFV}\approx \left(2\times 10^{-20}\,,\,2\times 10^{-14} \right)\left[{\left(\bar Y_N \bar Y_N^\dagger \right)_{12}\over 0.02,0.3}\right]^2\left(f_L\over 0.13,0.016\right)^4  \left({3\,{\rm TeV}\over M_{KK}}\right)^4\,,
\ee
 which is translated to the following bound on the Yukawa
\be
\left(\bar Y_N \bar Y_N^\dagger \right)_{12}\lsim(2,10^2)\times \left({3\,{\rm TeV}\over M_{KK}}\right)^2 \times \left({f_L\over 0.13,0.016}\right)^2\,,
\label{YNeee}
\ee
which is at present weaker than the bound from $\mu\to e$ conversion.

\subsubsection{Chiral-preserving UV Sensitive Contributions}
In this part we discuss additional one loop divergent contributions relevant to the above observables. For simplicity
we discuss only the Higgs on the IR brane case where the resulting constraints are more severe since the degree of divergence is higher.
There are two types of divergences: The first quadratic and it is due to the fact that the Yukawa couplings are corrected at the one loop level. The second is linear and it is due to one loop
divergent contribution to the IR localized brane kinetic terms (BKTs). We start with the former, it is of the form:
\be
\bar Y_E^{(\rm loop)} \sim {\Lambda^2_Y\over k^2 16 \pi^2}  \bar{Y}_N \bar Y_N^\dagger Y_E
\approx {N_{KK}^2\over 16 \pi^2}  \bar{Y}_N \bar Y_N^\dagger Y_E\,,
\ee
where in the above we have kept only the leading flavor violating contributions which are quadratically divergent and $\Lambda_Y$ stands for the cutoff on the IR brane.
The second contribution involves brane kinetic term for the LH fields
\be
\left(\bar L D \hspace*{-.25cm}\slash L\right)^{(\rm loop)}_{IR}\sim \bar L D \hspace*{-.25cm}  \slash \, {\Lambda_Y\over k 16 \pi^2}  \bar{Y}_N \bar Y_N^\dagger\, L\,.,
\ee
 Note that by gauge invariance we can relate this contribution to the coupling of the KK $Z$
 to the LH leptons, which mixes with the physical $Z$ due to EWSB.
 
These corrections spoil the perfect alignment between the charge lepton couplings to the physical $Z$.
Note that the first contribution is relevant only when the couplings to the physical $Z$ are non-universal.
Therefore, only the singlet charge leptons are affected by this misalignment.
This implies that the the amplitude for transition between the $i$th and $j$th  generation is suppressed by $f_{E^i}/f_{E^j}$.
The bound from the $\mu\to e$ transition is obtained by replacing $g_{L}^{\mu e}$  with $ g_{R}^{\mu e}$
in Eq.~({\ref{coeff}}) where
\be
g_{R}^{\mu e}&\approx&  f_{E^3}^2\, { m_e m_\mu \over m_\tau^2}  \,(\bar Y_N \bar Y_N^\dagger)_{12}
 {N_{KK}^2\over 16 \pi^2} \, k\pi r_c \left( {M_Z\over  M_{KK}}\right)^2
\,. \label{coeffR} 
\ee
We find that 
\be
g_{R}^{\mu e}/g_{L}^{\mu e}\sim  {f_{E^3}^2\over f_L^2}\,{ m_e m_\mu \over m_\tau^2}   \, {N_{KK}^2 k\pi r_c\over 2\ln \left( {M_W^2\over M_{KK}^2}\right) }\,,
\ee
which implies that this contribution is negligibly small over the all allowed parameter space (assuming $N_{KK},M_{KK}=3,3\,$TeV respectively). Clearly the corresponding contribution to other flavor violating processes are more suppressed.

The contribution of the second kind to $\mu\to e$ conversion yield the following
effective coupling
\be
\left(g_{L}^{\mu e}\right)^{KK}&\approx&  f_{L}^2  \,(\bar Y_N \bar Y_N^\dagger)_{12}
 {N_{KK}\over 16 \pi^2} \, k\pi r_c \left( {M_Z\over  M_{KK}}\right)^2
\,, \label{coeffLKK} 
\ee
we find that 
\be
\left(g_{L}^{\mu e}\right)^{KK}/g_{L}^{\mu e}\sim   {N_{KK} k\pi r_c\over 2\ln \left( {M_W^2\over M_{KK}^2}\right) }\sim 10\times {N_{KK}\over 3}\,.
\ee
This can be translated to the following bound on the neutrino Yukawa coupling:
\be
\left(\bar  Y_N \bar Y_N^\dagger \right)_{12}\lsim(0.07,5)\times \left({3\,{\rm TeV}\over M_{KK}}\right)^2 \times {N_{KK}\over 3}\times \left({f_L\over 0.13,0.016}\right)^2\,,
\label{YNconv}
\ee
still weaker than the bounds obtained via $\mu\to e\gamma\,,$ in particular when $f_L$ is set to its minimal
value.
The constraints for the other processes are scaled precisely  in the same way and are therefore weaker.


\subsection{ $\tau$ Decay Predictions}
We next focus on the model prediction involving $\tau$ flavor violation.
The Lagrangian for $\tau\to\mu$ decay is obtained from Eq. (\ref{Leff}) by replacing $\mu$ with $\tau$ and $\mu$ with $e$.
We find the following prediction for the various leptonic final states~\cite{PDG}:
\begin{eqnarray}
BR(\tau \rightarrow \mu \gamma)&\sim& 4.8 \times 10^{-8} \times \left(\bar{Y}_N \bar Y_N^\dagger\right)^2_{23} \times \left({3\,{\rm TeV} \over M_{KK}} \right)^4 BR(\tau \to e\nu\nu)< 3.2\times 10^{-8},  \\
BR(\tau \rightarrow 3\mu) &\sim& \left(2\times 10^{-20}\,,\,2\times 10^{-14} \right)\left[{\left(\bar Y_N \bar Y_N^\dagger \right)_{23}\over 0.02,0.3}\right]^2\left(f_L\over 0.13,0.016\right)^4  \left({3\,{\rm TeV}\over M_{KK}}\right)^4 BR(\tau \to e\nu\nu)<3.2\times 10^{-8}, \nonumber \\
BR(\tau \rightarrow 3e) &\sim&  \left(2\times 10^{-20}\,,\,2\times 10^{-14} \right)\left[{\left(\bar Y_N \bar Y_N^\dagger \right)_{13}\over 0.02,0.3}\right]^2\left(f_L\over 0.13,0.016\right)^4  \left({3\,{\rm TeV}\over M_{KK}}\right)^4 BR(\tau \to e\nu\nu)<3.6\times 10^{-8},\nonumber \\
BR(\tau \rightarrow \mu ee) &\sim& \left(1\times 10^{-20}\,,\,1\times 10^{-14} \right)\left[{\left(\bar Y_N \bar Y_N^\dagger \right)_{23}\over 0.02,0.3}\right]^2\left(f_L\over 0.13,0.016\right)^4  \left({3\,{\rm TeV}\over M_{KK}}\right)^4 BR(\tau \to e\nu\nu)<2.7\times 10^{-8},\nonumber \\
BR(\tau \rightarrow e\mu\mu) &\sim&  \left(1\times 10^{-20}\,,\,1\times 10^{-14} \right)\left[{\left(\bar Y_N \bar Y_N^\dagger \right)_{13}\over 0.02,0.3}\right]^2\left(f_L\over 0.13,0.016\right)^4  \left({3\,{\rm TeV}\over M_{KK}}\right)^4  BR(\tau \to e\nu\nu)<4.1\times 10^{-8}. \nonumber
\label{bfracs}
\end{eqnarray}
where $ BR(\tau \to e\nu\nu)\simeq 0.18 $~\cite{PDG}.

In an anarchic setup where all the above coefficients are determined from $Y_N$s that all have comparable values, we expect the rates to be comparable to those for the muon processes (up to a factor of $ BR(\tau \to e\nu\nu)\simeq 0.18$). Given  the current strong muon decay bounds that constrain $Y_N$ to be of order 0.1, we don't expect these rates to be measurable unless there is a sizable enhancement.
Note that this is true even for the two-body decay since the operator coefficient through which the decay occurs is proportional to $m_\tau$ so the combination of phase space and squared amplitude has the same $\tau$ mass dependence. We now discuss possibilities for tau decays consistent with our anarchy assumption.

In the more symmetric case when the $U(3)_L\times U(3)_N$ flavor symmetry is broken only on the IR brane and the right-handed neutrinos are degenerate and the neutrino mass hierarchy is accounted for solely through the Yukawa the model has
 extra predictive power.  In this case the above contributions to flavor violation processes are proportional
to $Y_N Y_N^\dagger$ in the mass basis for the charged lepton.
In this basis $Y_N Y_N^\dagger=V_{\rm MNS}Y_N Y_N^\dagger V_{\rm MNS}^\dagger\,,$
where the elements relevant to flavor violation are approximately
\be
(V_{\rm MNS} Y_N Y_N^\dagger V_{\rm MNS}^\dagger)^{\rm normal}_{12,13,23} &\propto&
\sqrt{\Delta m^2_{sol}}\,C_{12} C_{23} S_{12}\,, \ -\sqrt{\Delta m^2_{sol}}\, C_{12} S_{12} S_{23}\,, \
\sqrt{\Delta m^2_{atm}} \, C_{23} S_{23}\,,
\ee
To get an idea of the size of these predictions we take for simplicity $S_{12,23,13}^2\sim1/3,1/2,0$. Here the ratio of tau flavor-changing decays to muons to muon flavor-changing decays to electrons is of order
\be
{BR(\tau \to \mu) \over BR(\mu \to e) } \sim \left ({3 \over 2} \right)^2 
\left( {\Delta m^2_{atm} \over \Delta m^2_{sol} } \right) / BR(\tau \to e\nu\nu)\,.
\ee

where $S_{ij}\equiv \sin\theta_{ij}\,,$  $C_{ij}\equiv \cos\theta_{ij}\,,$ and  corrections of $\ord{\sqrt{\Delta m^2_{sol}/\Delta m^2_{atm}}}$ have been neglected. 
The maximal value for this ratio is about 10 and will be very difficult to measure in any tau decay process given the current constraints on electron processes.

The only case that has a chance of being measurable is the somewhat less generic inverted hierarchy case 
\be
(V_{\rm MNS} Y_N Y_N^\dagger V_{\rm MNS}^\dagger)^{\rm inverted}_{12,13,23} &\propto&
{\Delta m^2_{sol} \over 2 \sqrt{\Delta m^2_{atm}}}\,S_{12} C_{12} C_{23} \,, 
\ {\Delta m^2_{sol} \over 2 \sqrt{\Delta m^2_{atm}}}\,S_{12} C_{12} S_{23} \,,
 \ \sqrt{\Delta m^2_{atm}} \, S_{23}  C_{23}\,;
\ee
where the above ratio is about 
\be
{BR(\tau \to \mu) \over BR(\mu \to e) } 
\sim 
9
\left({\Delta m^2_{atm} \over \Delta m^2_{sol} }\right)^2
 / BR(\tau \to e\nu\nu)\,,
\ee
which can be about 1000 and might be measurable at future $b$-$\tau$ factories.

Note that in all cases the $\tau \to e$ rates are comparable to the $\mu \to e$ rates.
Note these results are rather general and do not necessarily distinguish our type of model in which flavor-changing in the charged lepton sector corresponds to minimal flavor violation and the dominant contributions are through intermediate states related to neutrinos so that it depends on the neutrino Yukawa matrix. Such a contribution is present in supersymmetric modesl as well for example. Tau decay predictions will be larger only in models where some other contribution dominates in which the $\mu \to e$ processes are suppressed. Such examples include GUT models where mixing feeds in from the quarks and the dominant contributions involve the third generation. In such a case the $\mu$ decay is smaller so that $\tau$ decays might be visible even with constraints on the muon sector. However, in general, there is no reason for the neutrino Yukawa to be suppressed.

\subsection{Constraints from Flavor Diagonal Processes}
We finish by discussion the model's contribution to the anomalous magnetic moment of the muon, $g-2$,
and the electron electric dipole moment (EDM), $d_e$. Both are UV sensitive with an IR brane Higgs.
In our model the leading contributions to $g-2$ are due to the same diagram that induces $\mu\to e\gamma\,.$
As we have discussed, due to anarchy in general there is no extra suppression associated with lepton flavor conversion.
It is rather easy to see why this is the case:
We can invert the relation in Eq.~(\ref{mueg}) to find that $|C|\lsim 10^{-14}\,.$
Where as $\Delta a_\mu$, the contributions to $g-2$, is given by $C/e\lsim 10^{-13}$ 
which is more than four order of magnitudes smaller  the SM  EW contributions. Thus the contributions to $g-2$ should be beyond the reach of experiments and cannot account
for the roughly 3 $\sigma$ observed deviation from the SM predictions~\cite{g-2}. 

This is roughly in agreement with naive dimensional analysis since
the ratio between the SM EW contribution and the ones from our model is roughly 
$(M_W/M_{KK})^2\,\bar Y_N^2\sim 10^{-4}$ when setting $M_{KK}=3\,$TeV and $\bar Y_N^2\sim (0.03,0.2)$
according to Eq. (\ref{UVbound}).

\begin{figure}[htbp] 
   \centering
   \includegraphics[width=2.5in,angle=270]{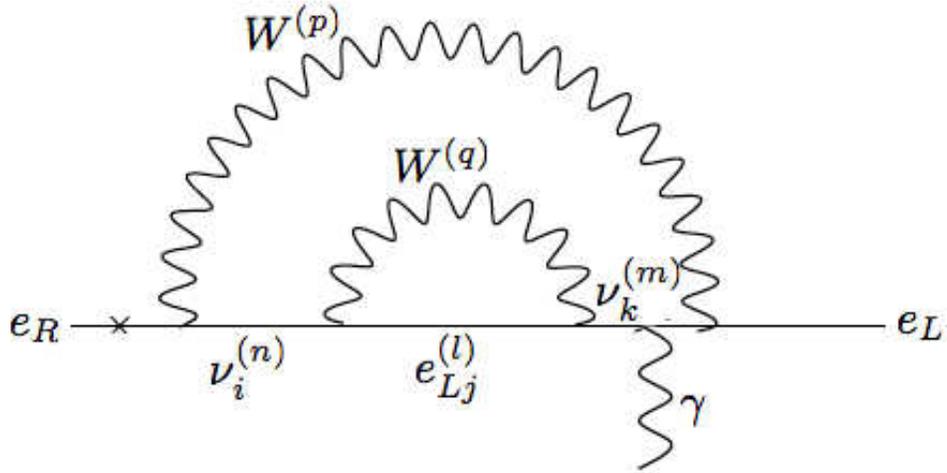} 
   \caption{An example for one of the leading contributions to the electron EDM, $d_e$.}
   \label{diagram3}
\end{figure}

As in~\cite{FPR} the one loop contributions to the electron EDM vanishes since due to LMFV they have to 
be proportional to the commutator between $Y_E$ and $Y_N$ (similar to the quark case, see {\it e.g}~\cite{BNS}).
In fig. \ref{diagram3} we show one of the 2-loop diagrams that yield the leading contributions for illustration. 
Here we do not aim towards providing the full calculation of the corresponding contributions (which in the brane Higgs case are  UV sensitive). However to get a rough bound we can estimate the contribution due to the diagram in the figure.
The KK masses square in the loop are roughly $(M_{KK}^{\nu,e})^2=  M_{KK}^2 +v^2 \bar Y_{N,E}\bar Y_{N,E}^\dagger$
where $M_{KK}\sim 3\,$TeV. Thus we estimate the electron EDM by the following
\be
\hspace*{-.2cm}
d_e &\sim& (k\pi r_c)^2 {N_{KK}^2 \,e \,m_e v^6 \over  (16 \pi^2 M_{KK} )^8 }\times V_{\rm MNS}^{1i} f(x^{N}_{i}) V_{\rm MNS}^{ij} f(x^{E}_{j}) V_{\rm MNS}^{jk} f(x^{N}_{k}) 
V_{\rm MNS}^{k1}  \nonumber \\ 
&\sim&   (10^{Ð37},\,10^{-31}) \left(1,\,{\Delta m_{sol}^2\over \Delta m_{atm}^2}\right) J^{MNS}   
 \left({\bar Y_N \bar Y_N^\dagger \over 0.02,0.3}\right)^2
 \left({ \bar Y_E \bar Y_E^\dagger \over 0.2,4\times 10 } \right)
 \times \left({3\,{\rm TeV}\over M_{KK}} \times {N_{KK}\over3}\right)^2  {\rm e \, cm},
\ee
where $x^{E,N}_i= {\rm diag}(\bar Y_{E,N})^2 $
 and $f(x^{N,E}_{i})$ is a function of differences of the $x^{N,E}_i$s
due to GIM.  
The factor $ \left(\Delta m_{sol}^2\over \Delta m_{atm}^2\right) J^{\rm MNS}$ on the RHS of the relation is  due to GIM suppression 
 and $J^{\rm MNS}\sim S_{13}< 0.1$
is the $V_{\rm MNS}$ Jarlskog invariant setting the possible Majorana phases to zero~\cite{Jarlskog} 
As before we have used the extreme values for $\bar Y_N$ given in Eqs. 
 (\ref{UVbound},\ref{1loopbound}) to get the upper bound on $d_e$ in the case of an IR Higgs and bulk Higgs respectively. 
To find the corresponding extremes for $Y_E $ we use the bound in Eq.~(\ref{YEbound}) 
for its lowest possible value (corresponding to a brane Higgs) while the maximal value is
taken to saturate the strong coupling upper bound in the case of a bulk Higgs~\cite{FPR} and the suppression proportional to $ \left(\Delta m_{sol}^2\over \Delta m_{atm}^2\right)$  is for degenerate right-handed neutrinos.
We see that the current experimental bound, $d_e=(7 \pm 7 )\,10^{-28}\ {\rm e \, cm}$~\cite{PDG} is far from the model range of prediction
however near future proposed experiments  might probe the relevant parameter space~\cite{eEDMexp}.



\section{Conclusion}

We have shown how lepton  masses and mixings fit in nicely into a warped geometry framework where wavefunctions derived from bulk masses account for both. We have seen the range of masses and mixings can be accounted for with no extremely large or small parameters and that furthermore the limited range of masses can be accounted for with LMFV, in which the masses are SU(3) symmetric broken only by Yukawas. Furthermore, with QMFV and LMFV, the difference between small quark mixing angles and large neutrino mixings is explained by  more degenerate left-handed lepton wavefunctions.

Our model has several predictions, though no true smoking gun. Soft imprecise predictions include the form of the neutrino mass matrix. We expect that mixing angles to be of order unity but not a precise angle derived from group theory or some underlying precisely predicted model consideration. If mixings are of order unity to within a percent, we would expect our model to be less likely. Similarly, if $\theta_{13}$ turns out to be zero to within a percent (again this is a loose consideration) we would expect our model is not correct. We predict generic order unity mixings but no precise numerical values. 

Predictions for flavor-changing processes depend on the right-handed neutrino masses, which, along with measured masses and mixings, determine the Yukawas on which the mixing predictions depend. We have seen that in general flavor-changing $\tau$ decays will not be observable unless neutrinos appear in an inverted hierarchy case which is less favored in the most natural setting. Certainly observing tau flavor-changing decays in the absence of an inverted hierarchy would rule out our model.

Our model makes sense in the context of Majorana neutrino masses. We predict  standard neutrinoless beta decay and CP violation in the neutrino sector.
Furthermore, our model relies on matter and gauge bosons in the bulk. Therefore it would predict the bulk KK gauge bosons considered in~\cite{FPR,gaugeflavor,shiningRS}.
The phenomenology of this new sector is very interesting and deserve a detailed study (which is beyond the scope of this work)~\cite{shiningRS}.
However, the point is that as long as the symmetry breaking is consistent with LMFV then we still expect, leading order, alignment between
the new sources of lepton flavor breaking and lepton masses which does not induce flavor violation.
In terms of 4D global symmetries of the lepton sector, $SU(3)_L \times SU(3)_E\times U(1)_{\rm L}$, our model 
utilizes the split fermion mechanism to breaks $SU(3)_E\times U(1)_{\rm L}$ in a conventional way. The breaking of 
$SU(3)_L$ is inspired by dynamical breaking (occur dominantly in the IR) which can be understood via the AdS/CFT correspondence, and
LMFV and RS-GIM provide protection against the extra sources of lepton flavor violation. 
The fact that in our model the SM approximated global currents are gauged~\cite{APR,cust1} is also motivated by the correspondence and we
find it therefore appealing that following the AdS/CFT prescription the model is become
more consistent with the data.

Finally our model can be made to be consistent with unification (for a unification within the RS framework see 
{\it e.g} \cite{GUT} and Refs. therein). We find it remarkable that the large range of masses and mixings in the quark and lepton sector can be explained with no unusually large or small parameters in an economical explanatory framework. Hopefully experiments will test these ideas in the near future.

\vspace*{1cm}


{\bf Acknowledgements:} We thank Kaustubh Agashe, Liam Fitzpatrick, Concepcion Gonzalez-Garcia, Yuval Grossman, and Raman Sundrum for useful discussions.
The research of GP is supported by and NSF grant PHY-06353354.
LR is supported by NSF grants PHY-0201124 and PHY-055611.  
GP  thanks the hospitality of the theoretical physics group of Boston and Harvard universities
where part of this work was done.
LR thanks NYU and the NYU Physics Department and CCPP
for their kind hospitality and support and also the Caltech Moore Scholarship program. This work was completed while LR was a Moore
Distinguished Scholar at Caltech.

Note added: while this work near completion Ref. \cite{Chen:2008qg}
was published which also deals with the RS lepton flavor problem via 5D MFV.
However, the model presented requires introducing the neutrino anarchy by hand and also rely on
bulk global U(1) lepton symmetry.

\end{document}